\documentclass[prd,twocolumn,preprintnumbers]{revtex4}

\usepackage{amsmath}
\usepackage{epsfig}
\usepackage{graphicx}
\usepackage{color}
\usepackage[normalem]{ulem}
 \usepackage{url}
\usepackage[breaklinks, plainpages=false, colorlinks=true, anchorcolor=cyan, linkcolor=red, citecolor=cyan, urlcolor=magenta, bookmarks=false]{hyperref}

\usepackage[caption=false]{subfig}

\begin{document}
\renewcommand{\thefigure}{\arabic{figure}}
\setcounter{figure}{0}

 \def\I{{\rm i}}
 \def\E{{\rm e}}
 \def\D{{\rm d}}

\bibliographystyle{apsrev}

\title{Heterodyned Likelihood for Rapid Gravitational Wave Parameter Inference}

\author{Neil J. Cornish}
\affiliation{eXtreme Gravity Institute, Department of Physics, Montana State University, Bozeman, Montana 59717, USA}

\begin{abstract} 

Inferring the source properties of a gravitational wave signal has traditionally been very computationally intensive and time consuming. In recent years, several techniques have been developed that can significantly reduce the computational cost while delivering rapid and accurate parameter inference. One of the most powerful of these techniques is the heterodyned likelihood, which uses a reference waveform to base-band the likelihood calculation. Here an efficient implementation of the heterodyned likelihood is presented that can be used for a wide range of signal types and for both ground based and space based interferometers. The computational savings relative to direct calculation of the likelihood vary between two and four orders of magnitude depending on the system. The savings are greatest for low mass systems such as neutron star binaries. The heterodyning procedure can incorporate marginalization over calibration uncertainties and the noise power spectrum.
\end{abstract}

\maketitle

\section{Introduction}

Full parameter inference~\cite{Veitch_2015,Ashton_2019} for gravitational wave signals can be very computationally intensive, with runs on single systems taking days or weeks using contemporary hardware. This inefficiency has motivated the development of novel approaches to speed up the process. Some of these methods work by speeding up the calculation of waveform templates using singular value decomposition~\cite{Cannon:2011rj,Smith_2013} or reduced-order modeling~\cite{Field:2011mf,Field:2013cfa}. Other methods speed up the likelihood evaluation using techniques such as heterodyning~\cite{Cornish:2010kf,Cornish:2020vtw,Cornish:2021wxy,Zackay:2018qdy}; waveform decomposition and pre-computation~\cite{Pankow:2015cra,Cornish:2016pox,Cornish:2020vtw,Cornish:2021wxy}; reduced order quadrature~\cite{PhysRevD.94.044031,Morisaki:2020oqk}; and variable frequency banding~\cite{Morisaki:2021ngj}. Hardware based acceleration has also been investigated~\cite{Wysocki:2019grj,Smith:2019ucc,Katz:2021yft}, as has machine learning~\cite{George:2017pmj,Delaunoy:2020zcu,Dax:2021tsq}.

The heterodyning approach, first introduced in 2010~\cite{Cornish:2010kf}, and later re-branded under the unfortunate term ``relative binning''~\cite{Zackay:2018qdy}, has yet to be widely adopted, despite it being widely applicable, easy to implement and incredibly fast. The goal of this paper is to present an efficient implementation of the heterodyned likelihood using discrete Legendre polynomial expansions on dynamically spaced frequency grids. The speed and accuracy of the heterodyning approach is demonstrated, along with a discussion of how it can be used when marginalizing over calibration and noise models.

\section{The Heterodyned Likelihood}

The idea behind the heterodyned likelihood is very simple~\cite{Cornish:2010kf}. To match the signal well enough to give a decent likelihood, the phase and amplitude evolution of the waveform template has to closely match that of the signal. Thus if $\bar{h}(f)=\bar{\cal A}(f) e^{i\bar \Phi(f)}$ is a reference template with high likelihood and $h(f)={\cal A}(f) e^{i\Phi(f)}$ is another template with high likelihood, the ratio $\zeta(f) = h(f)/\bar{h}(f)$ will be a slowly varying function. For signals made up of multiple harmonics the same reason applies harmonic by harmonic. The terms that appear in the log likelihood,
\begin{equation}
\ln L=  (d|h) - \frac{1}{2} (h|h),
\end{equation}
where $d$ is the data and $(a|b)$ denotes the noise weighted inner product, can be factored into slowly varying and rapidly varying components:
\begin{eqnarray}
(d|h) &=& 2 \int \frac{ d(f) h^*(f) + d^*(f) h(f)}{S(f)} df \nonumber \\
&=& 2 \int (\kappa(f) \zeta^*(f) + \kappa^*(f) \zeta(f)) df
\end{eqnarray}
where $\kappa(f) = d(f) \bar{h}^*(f)/S(f)$ is rapidly varying and $\zeta(f)$ is slowly varying. Similarly, 
\begin{eqnarray}
(h|h) &=& 2 \int \frac{ h(f) h^*(f) + h^*(f) h(f)}{S(f)} df \nonumber \\
&=& 4 \int  df |\zeta(f)|^2 \sigma(f) df
\end{eqnarray}
where $\sigma^2(f) = |\bar{h}(f)|^2/S(f)$ is rapidly varying and $|\zeta(f)|^2$ is slowly varying (the rapid variation of $\sigma(f)$ is due to spectral lines). The heterodyning procedure uses a low order polynomial expansion of the slow terms to dramatically decrease the computational cost of computing the likelihood. The accuracy of the approximation can be improved by writing the likelihood as
\begin{equation}
\ln L=   \ln \bar{L} -  (\bar{r} |\Delta h) - \frac{1}{2} (\Delta h| \Delta h),
\end{equation}
where $ \ln \bar{L}$ is the likelihood computed using the reference waveform, $\bar{r} = d - \bar{h}$ is the reference residual, and $\Delta h = \bar{h} - h$. Since the terms involving $\Delta h$ are generally small, the fractional error introduce by approximating the integrals results in a smaller absolute error than approximating the full integrals $(d|h)$ and $(h|h)$. The slow varying term is now $\Delta \zeta(f) = \Delta h(f)/\bar{h}(f) = 1- \zeta(f)$.

\subsection{Legendre Expansion}

In practice, the integrals become sums over frequency that can be efficiently computed using an expansion in discrete Legendre polynomials $P_{n k}$ where $n$ denotes the polynomial order and $k=0,1,..,N$ denotes the frequency bin. The $P_{n k}$ satisfy the orthogonality condition
\begin{equation}
P_{nk} P_{mk} = \delta_{nm} \alpha_{nN} \, ,
\end{equation}
where $\alpha_{nN}$ is a normalization factor~\cite{legendre} and we have used the Einstein summation convention. The first few $P_{n k}$ have the form
\begin{eqnarray}
P_{0k} &=& 1 \nonumber \\
P_{1k} &=& 1 - \frac{2 k}{N} \nonumber \\
P_{2k} &=& 1 - \frac{6 k}{N} +  \frac{6 k (k-1)}{N (N-1)} \, .
\end{eqnarray}
Note that the shape of the polynomials depends on the number of frequency bins, $N+1$. The higher order polynomials can be generated efficiently using a recursion relation~\cite{legendre}.

A function $g_k = g(f_k)$ can be expanded:
\begin{equation}
g_{k} = \gamma_n  P_{n k}\, ,
\end{equation}
where $n=0,1,..N$ and 
\begin{equation}\label{coeff}
\gamma_n  =   \frac{P_{n k} g_k}{\alpha_{nN}}\, .
\end{equation}
In this way, a term in the likelihood such as $(d|h)$ can be written as
\begin{equation}\label{expand}
(d|h) = 2 \alpha_{nN} (\kappa^R_{n} \zeta^R_n + \kappa^I_n \zeta^I_n)
\end{equation}
where, for example, $\{\kappa^R_{n}, \kappa^I_{n} \},$ denote the expansion coefficients of the real and imaginary parts of $\kappa(f)$.  The expression (\ref{expand}) is exact, it is simply a re-writing of the original sum.  As it stands, (\ref{expand}), actually represents an {\em increase} in the computational cost from ${\cal O}(N)$ to ${\cal O}(N^2)$ (though there are fast Chebyshev--Legendre transforms~\cite{doi:10.1137/130932223} that reduces the cost to ${\cal O}(N (\log N)^2/ \log\log N)$). The savings come by restricting the sum over $n$ in (\ref{expand}) to a small number of coefficients. This can be done with little loss of accuracy since the high order terms in the expansion of the slowly varying components diminish very quickly with increasing $n$. 

The sum (\ref{expand}) can be most efficiently approximated by breaking the full sum into smaller segments. The sum can be restricted to frequencies where the reference waveform contributes significantly to the signal-to-noise. The efficiency is further improved by dynamically determining the width of the segments such that the slow terms can be accurately covered by polynomials of some chosen order $J$. The coefficients of the fast varying terms have to be computed at the full frequency resolution, but this can be done once and the results stored. Thus, the cost of the heterodyned likelihood comes down to computing the coefficients of the slow components in each frequency band. This cost scales as $J^2 Q$, where $Q$ are the number of bands. The number of waveform evaluations for the slow terms scales as $M=J Q$. While increasing $J$ allows us to use wider bands for a given error tolerance, the increase in bandwidth scales slower than the polynomial order $J$, making it more efficient to use low polynomial order and more bands.

In a given frequency band covering $N+1$ frequency bins, the coefficients $s_n$ of the slowly varying terms $s(f_j)$ can be computed using a sub-sample of frequency bins. Since the discrete Legendre polynomials are not orthogonal on the sub-grid, the usual expression for the expansion coefficients (\ref{coeff}) can not be used. Instead we start with the defining relation
\begin{equation}\label{hfit}
s(f_j) = s_n  P_{n}(f_j) \equiv s_n  Y_{n j} \, ,
\end{equation}
where $Y_{n j} = P_{n}(f_j)$ is a $(J+1)\times(J+1)$ matrix containing the sub-sampled values of the discrete Legendre polynomials. We can then solve for the expansion coefficients: 
\begin{equation}
s_n = Y_{n j}^{-1} s(f_j)   \, .
\end{equation}
The frequency samples $f_j$ do not have to be evenly spaced, but the matrix $Y_{n j}$ can become ill-conditioned if the spacing of any two bins exceeds $\sim  2 N/J$. To ensure optimal accuracy it is best to space the sub-samples uniformly across each frequency band. The matrix inverses for each frequency band can be computed once and stored for later use.

\subsection{Frequency Spacing}

The goal is to resolve the slow varying function $\Delta \zeta(f)$ to sufficient accuracy using the smallest number of frequency samples. The first thing to consider is how far from the reference waveform any template is likely to be. For Gaussian posterior distributions, we know that twice the log likelihood is chi-squared distributed with $D$ degrees of freedom, where $D$ is the number of parameters in the model. This scaling holds remarkable well even when the posterior distributions are non-Gaussian. Thus, to account for any waveforms that will contribute to the posterior distribution we need to cover deviations from the reference likelihood of order a few times the standard deviation $\sqrt{D}$. As a proxy we can use the chi-squared $\chi^2 = (\Delta h | \Delta h)$ as a measure of the deviation from the reference waveform, and so long as the frequency grid accurately covers departures as large as $\chi^2 \sim 20 \rightarrow 50$ the parameter estimation will be reliable.

\begin{figure}[htp]
\includegraphics[width=0.48\textwidth]{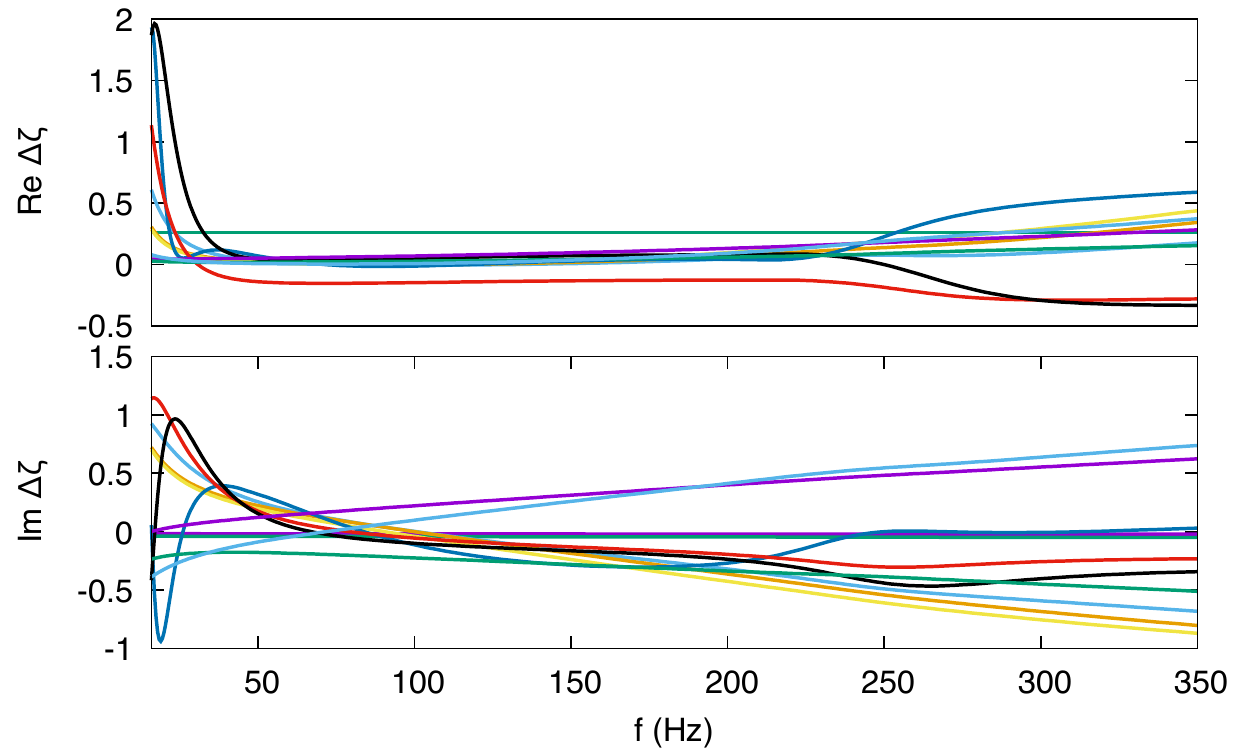} 
\caption{\label{fig:hetro} Real and imaginary parts of the heterodyned waveform differences, $\Delta \zeta = (\bar h - h)/\bar h$, with $h$ displaced along successive eigendirections of the Fisher information matrix, with the differences scaled such that $\chi^2 = (\Delta h| \Delta h)=50$.}
\end{figure}

By perturbing any one source parameter by a suitable amount, it is possible to arrive at the desired chi-squared, but the perturbation in each parameter will lead to a different behavior for the
$\Delta \zeta(f)$. One way to cover all possibilities is to consider perturbations along the eigendirections, $\vec{v}_i$, of the Fisher information matrix $\Gamma_{ij} = (\partial_{\theta^i} \bar h | \partial_{\theta^j} \bar h)$, such that $\Delta \vec{\theta}_i = \alpha_i \vec{v}_i$ (no sum on the $i$). In principle, setting $\alpha_i = \beta /\sqrt{\lambda_i}$, where $\lambda_i$ is the eigenvalue corresponding to the $\vec{v}_i$ eigenvector, should yield $\chi^2 = \beta^2$, but the Fisher matrix is often ill-conditioned, and the $\beta$'s need to be iteratively adjusted to give the desired chi-square. Figure \ref{fig:hetro} shows the real and imaginary parts of $\Delta \zeta$ for the black hole binary GW150914~\cite{LIGOScientific:2016aoc} evaluated at the LIGO Hanford detector and using the IMRPhenomD waveform model~\cite{Santamaria:2010yb}. The parameters were perturb along each eigendirection of the Fisher matrix and scaled to give $\chi^2 = 50$. We see that functions vary most rapidly at low frequencies and near merger (at around 200-300 Hz).

\begin{figure}[htp]
\includegraphics[width=0.48\textwidth]{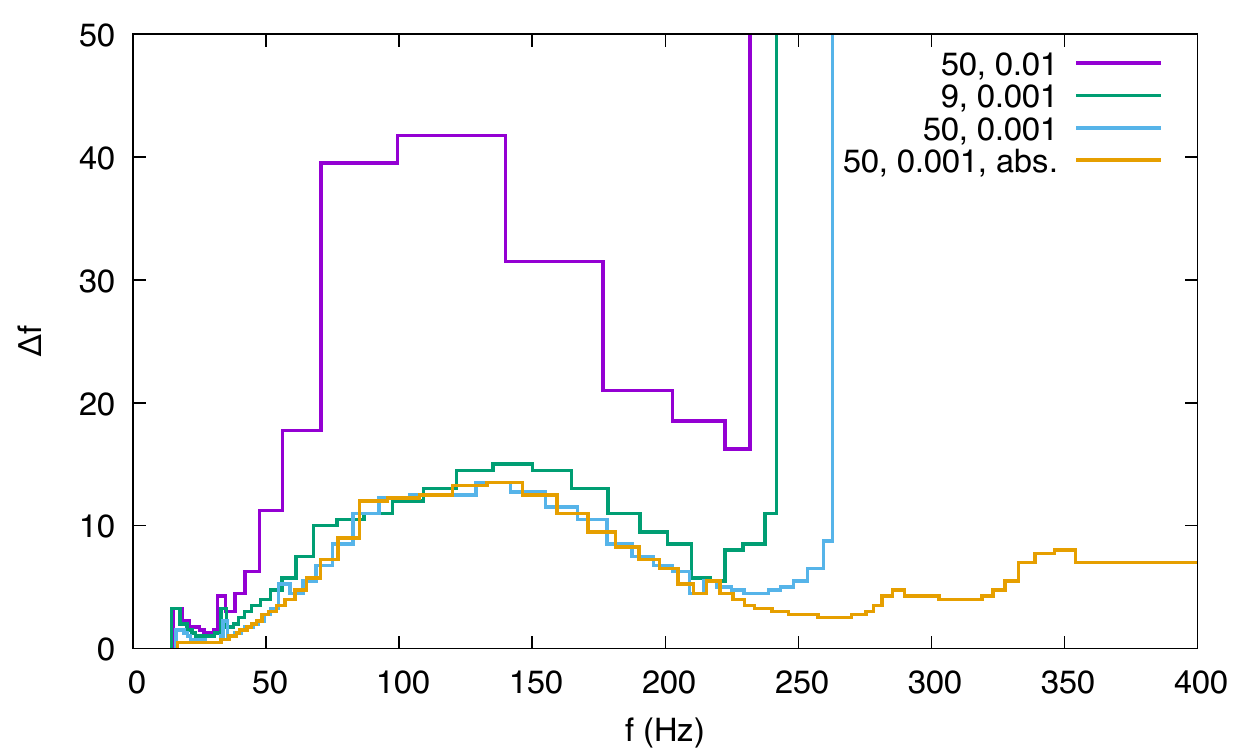} 
\caption{\label{fig:df1} Dynamic frequency spacing of the waveform samples for GW150914 using various choices of linear error tolerance and chi-squared offset. The spacings used a relative error tolerance, save for the one labeled ``abs'', which used an absolute error tolerance.}
\end{figure}

The frequency spacing can then be determined by considering a linear fit to the $\Delta \zeta$ for perturbations along each eigenvector direction, and in each detector. Starting with a frequency spacing of one frequency bin, the spacing is steadily incremented until the difference between $\Delta \zeta$ and the linear fit across the interval deviates by more than some specified tolerance. Figure~\ref{fig:df1} shows the frequency spacings for various error tolerances and chi-square values. We see that the samples are concentrated at low frequencies. To keep the number of samples to a minimum we can use a relative rather than absolute error tolerance. Since the contribution to the likelihood in each frequency band scales as $\sigma^2(f)=|\bar h(f)|^2/S(f)$, we can achieve a desired relative error tolerance by scaling the absolute error tolerance by $[\sigma^2]/\sigma^2(f)$ where the square brackets denotes the average value.

\begin{figure}[htp]
\includegraphics[width=0.48\textwidth]{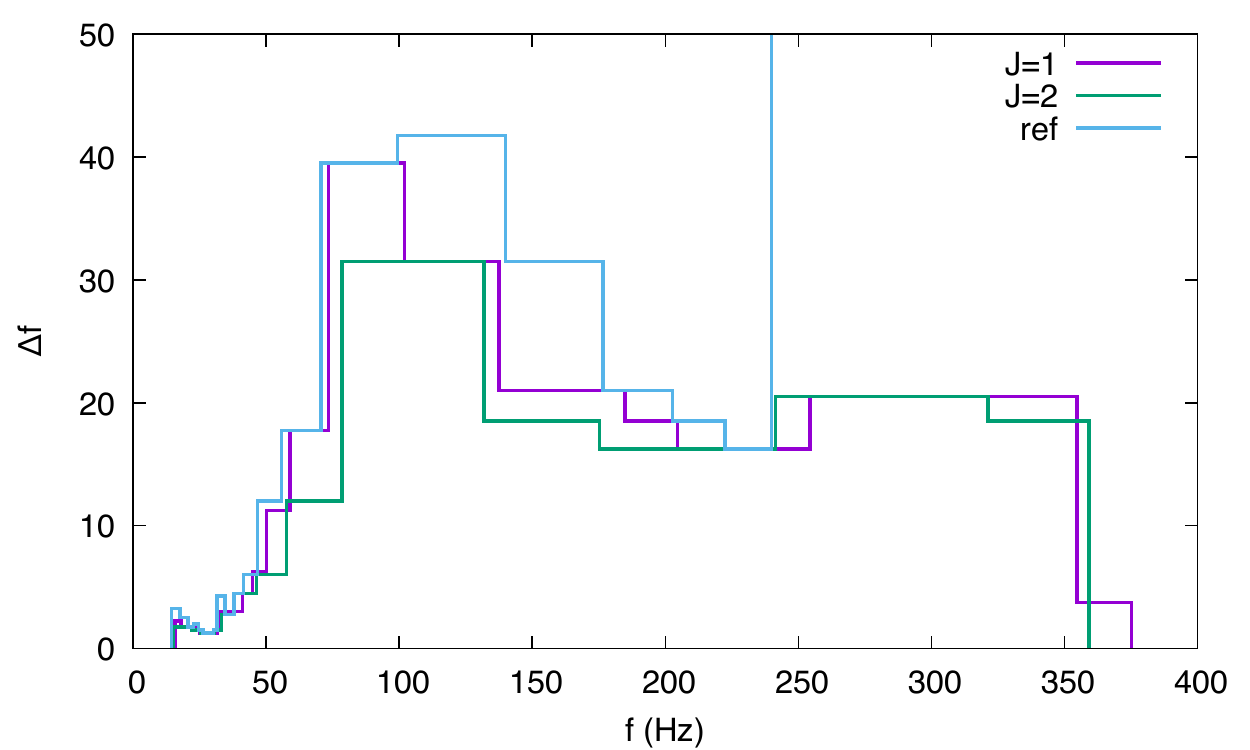} 
\caption{\label{fig:df2} Dynamic frequency spacing of the waveform samples for GW150914 at linear and quadratic order. The reference spacing is also shown.}
\end{figure}

The frequency bandwidths for higher order polynomial fits are derived from the reference linear fit by finding the smallest frequency spacing in a given region and multiplying that number by the polynomial order $J$. This results in slightly more samples than for the linear fit as the wider frequency bands are less able to accommodate to the ideal dynamic frequency spacing. Note that the enveloping procedure leads to slightly more samples being used even at linear ($J=1$) order. Figure \ref{fig:df2} shows the frequency spacing for GW150914 at linear and quadratic order compared to the reference spacing. There are 32 samples at linear order, 34 at quadratic order, as compared to 22 for the reference spacing. 

\begin{figure}[htp]
\includegraphics[width=0.48\textwidth]{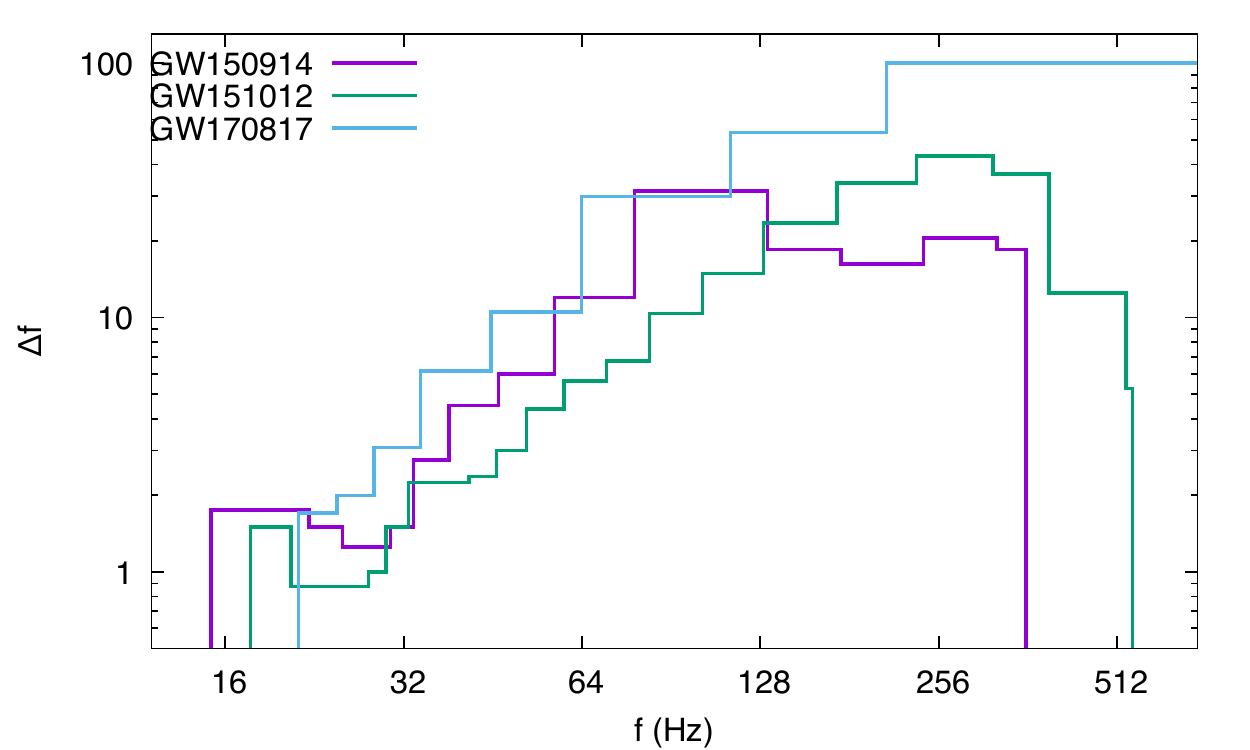} 
\caption{\label{fig:NSBH}  Comparison of the dynamic frequency spacing for the black hole binaries GW150914 and GW151012, and the binary neutron star binary GW170817. 
The heterodyne used second order Legendre polynomials with a linear fit tolerance of 0.01 and a chi-squared of $(\Delta h| \Delta h)=50$.}
\end{figure}

The frequency spacing dynamically adjusts to account for the characteristics of each system. Figure \ref{fig:NSBH} compares the frequency grids for three systems of decreasing total mass. In each case the minimum frequency was set at 8 Hz and maximum frequency was set at 1024 Hz. The time span analyzed increases with decreasing mass: 4 seconds for black hole binary GW150914, 8 seconds for black hole binary GW151012, and 128 seconds for the binary neutron star binary GW170817~\cite{LIGOScientific:2017vwq}. The number of frequency samples in the heterodyned likelihood were 34 for GW150914, 52 for GW151012 and just 22 for GW170817. The small number of samples for the neutron star binary GW170817 relative to the two black hole binaries is due to the fact that GW170817 entered the band at 22 Hz and exited the band prior to merger. The savings in waveform evaluations using the heterodyned likelihood relative to the direct likelihood the three systems are a factor of 120 for GW150914, 156 for GW151012 and 5900 for GW170817. 

\begin{figure}[htp]
\includegraphics[width=0.48\textwidth]{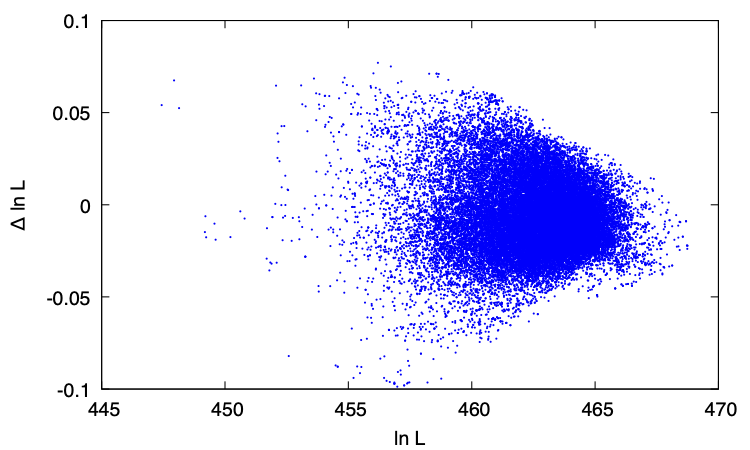} 
\caption{\label{fig:NSBHdL}  Difference between the full and heterodyned likelihood for neutron star binary GW170817. In both cases the heterodyne used second order Legendre polynomials with a linear fit tolerance of 0.01 and a chi-squared of $(\Delta h| \Delta h)=50$.}
\end{figure}

Figure \ref{fig:NSBHdL} shows the difference, $\Delta {\rm ln} L$ between the full and heterodyned likelihood calculation and a function of the likelihood for the samples collected during a Markov Chain Monte Carlo (MCMC) run on the GW170817 data. The absolute value of the error never exceeds 0.1, and the average absolute error is just $1.6 \times 10^{-2}$.

\begin{figure}[htp]
\includegraphics[width=0.48\textwidth]{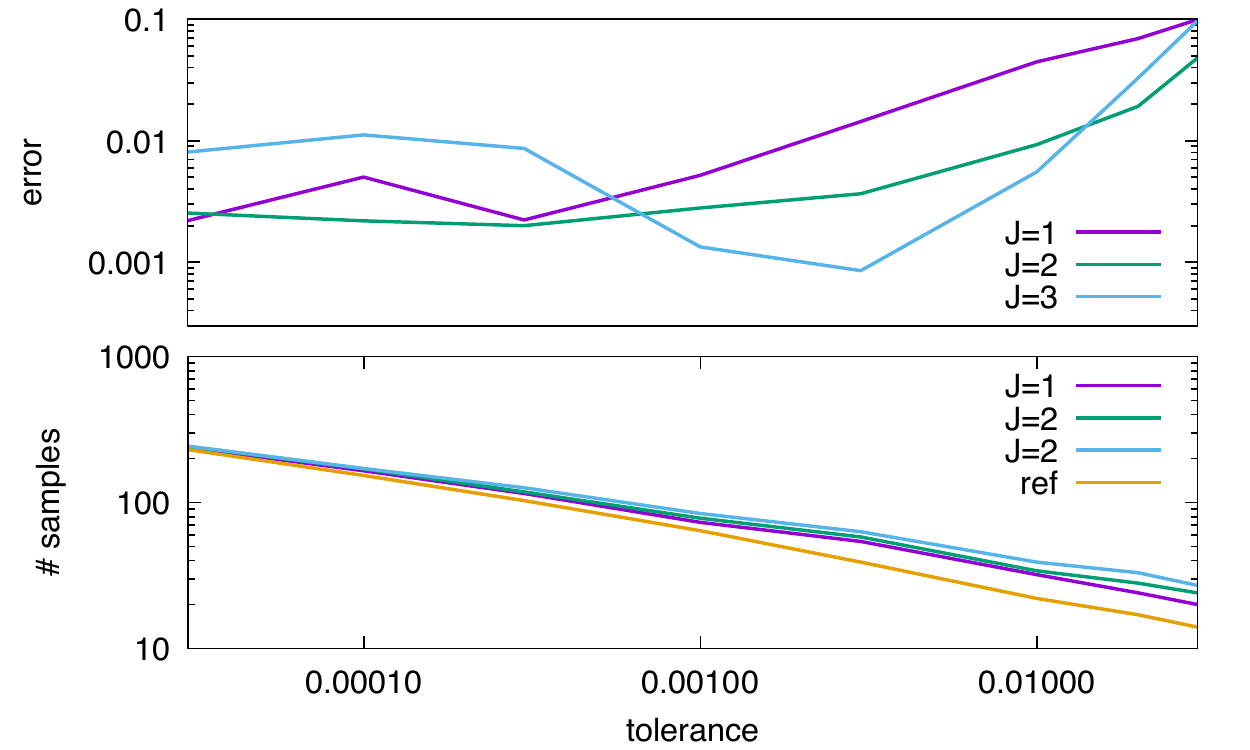} 
\caption{\label{fig:tol} The upper panel shows average of the absolute value of the difference between the full likelihood and the heterodyned likelihood as a function of the linear error tolerance for GW150914 at linear, quadratic and cubic order. The lower panel shows the number of frequency sample used in the heterodyne as a function of the linear error tolerance.}
\end{figure}

Smaller linear error tolerances result in more frequency samples in the heterodyne and a reduction in error in the likelihood. Figure \ref{fig:tol} shows how the error in the likelihood and the number of samples in the heterodyne scale with the linear error tolerance. The error in the likelihood is measured by the average of the absolute value of the difference between the full and heterodyned likelihood, $\overline{|\Delta {\rm ln} L|}$, for accepted samples in a MCMC run. The error decreases as the tolerance is reduced from 0.1 to 0.001, but then asymptotes or even increases. The reason for this behavior can be traced to the error introduced in estimating the Legendre expansion of the slow terms using (\ref{coeff}). As the error tolerance is decreased, the number of frequency bands grows as the error tolerance to the power $\sim -0.35$, leading to an increase in the number of error contributions. The errors in (\ref{coeff}) grow with polynomial order, as does the overall cost of computing the heterodyned likelihood. The sweet spot is to use a quadratic $(J=2)$ fit with an error tolerance between 0.01 and 0.001. While illustrated here for just one system, similar behavior was found to hold across the mass spectrum. 

\begin{figure}[htp]
\includegraphics[width=0.48\textwidth]{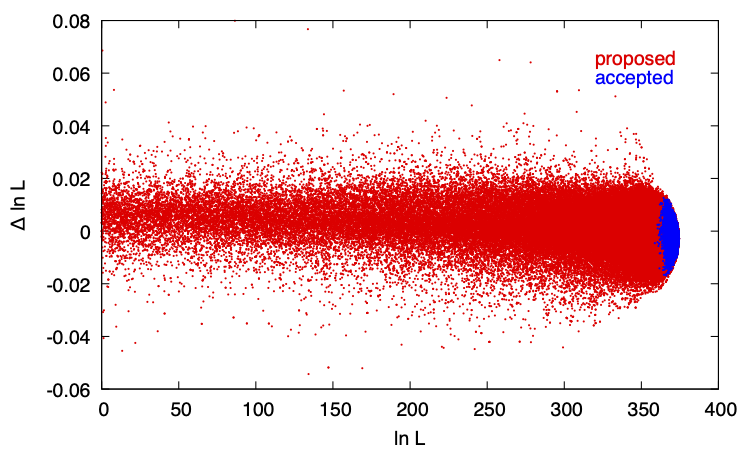} 
\caption{\label{fig:like} Difference between the full and heterodyned likelihood for GW150914 for proposed and accepted parameter values. The heterodyne used second order Legendre polynomials with a linear fit tolerance of 0.001 and a chi-squared of $(\Delta h| \Delta h)=50$.}
\end{figure}

If greater accuracy is desired, the number of samples in each frequency band used in the fit (\ref{hfit}) can be increased while keeping the polynomial order fixed, with the solution for the coefficients found using a singular value decomposition rather than the simple matrix inverse (\ref{coeff}). In considering what an acceptable error tolerance for $\Delta {\rm ln} L$ might be, recall that one standard deviation in the likelihood is $\sigma_{{\rm ln} L} \simeq  (D/2)^{1/2}$. For the IMRPhenomD waveform model~\cite{Santamaria:2010yb} used here, $D=11$ and $\sigma_{{\rm ln} L} \simeq  2.3$. The finite sampling that occurs in any numerical approach to Bayesian inference will introduce uncertainties of at least a few percent the posterior distribution, so demanding that $|\Delta {\rm ln} L| < 0.1$ should be sufficient for most applications. Interestingly, while the heterodyning procedure was designed to work for waveforms that are ``close'' to the reference waveform, it continues to work for waveforms that are far from the reference. This is illustrated in Figure \ref{fig:like}, where the error in the likelihood is show as a function for the likelihood for both accepted and proposed points in a MCMC run. Surprisingly, the heterodyne is accurate all the way down to zero log likelihood. This means that the heterodyned likelihood can not only be used for parameter estimation, but also for computing the model evidence using methods such as thermodynamic integration~\cite{Cornish:2007if}.

\begin{figure}[htp]
\includegraphics[width=0.48\textwidth]{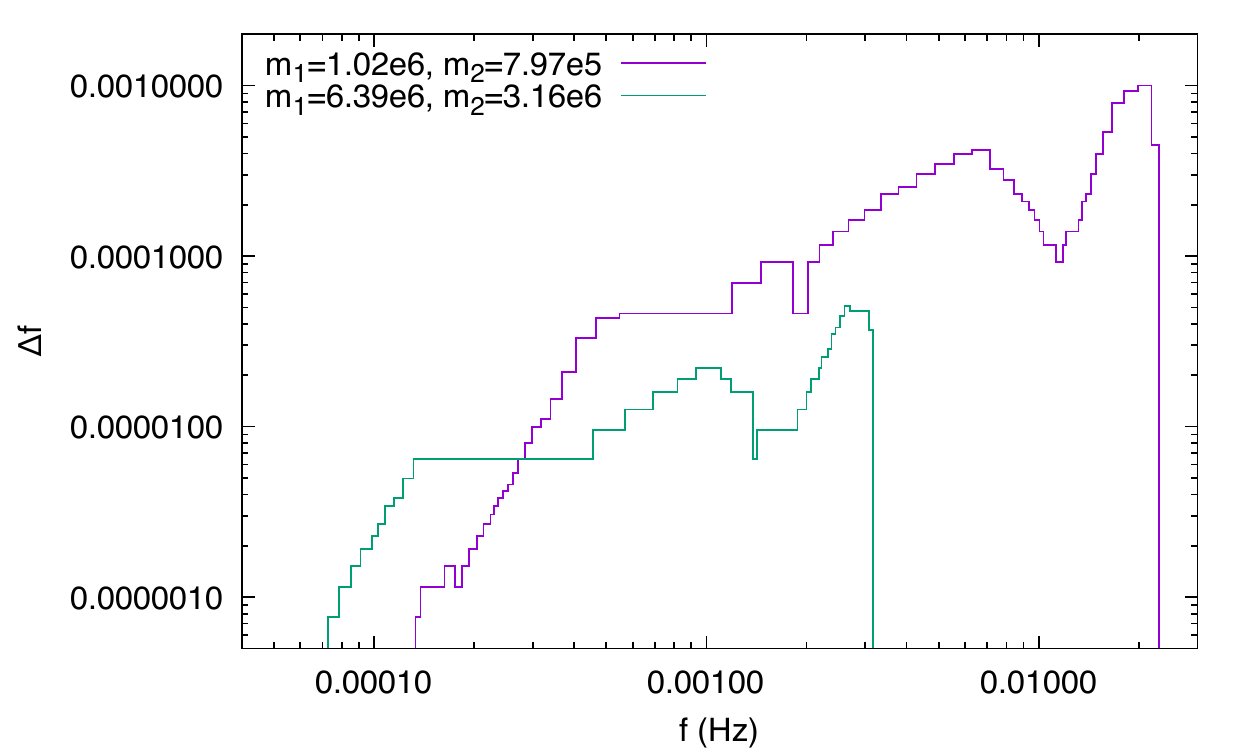} 
\caption{\label{fig:LISAdf}  Frequency spacing for two massive black hole binaries taken from the second LISA Data Challenge training data set. Despite the difference in total mass, both system required roughly two hundred frequency samples in the heterodyne. The detector frame masses of the black holes (in units of solar mass) are given in the figure legend. The heterodyne used second order Legendre polynomials with a linear fit tolerance of $10^{-5}$ and a chi-squared of $(\Delta h| \Delta h)=50$.}
\end{figure}

The heterodyning approach is widely applicable. For example, it can handle the very high signal-noise systems that are expected to be detected by the Laser Interferometer Space Antenna (LISA). The high SNR demands a smaller linear error tolerance, but the savings are still very large. For example, typical LISA sources with masses in the $10^5 \rightarrow 10^7 \, M_\odot$ range, such as those shown in Figure~\ref{fig:LISAdf}, can be accurately covered by frequency grids with just a few hundred points. This is far fewer than the $\sim 10^6$ samples that are required when computing the regular likelihood for these sources.

\begin{figure}[htp]
\includegraphics[width=0.48\textwidth]{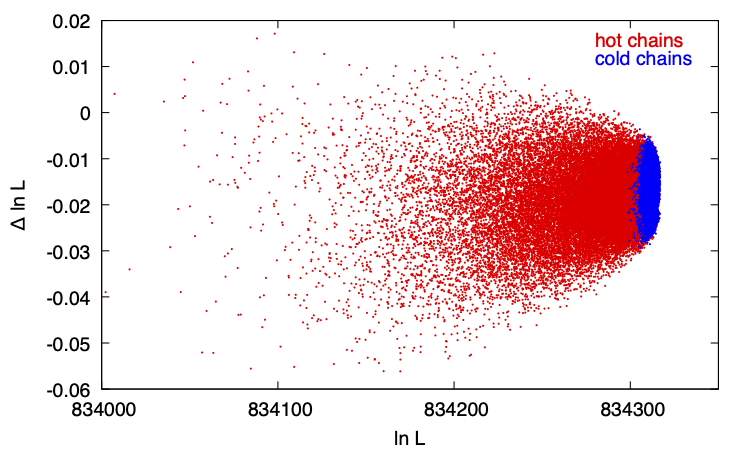} 
\caption{\label{fig:likeLISA} Difference between the full and heterodyned likelihood for a simulated LISA black hole binary with detector frame masses $m_1=1.02\times 10^6 \, M_\odot, \; m_2=7.97\times 10^5 \, M_\odot$. A parallel tempered MCMC was used with four cold chains and twelve hot chains, geometrically spaced in temperature by a factor of 1.3. The signal to noise of the system was very high: ${\rm SNR} = 1292$. The heterodyne used second order Legendre polynomials with a linear fit tolerance of $10^{-5}$ and a chi-squared of $(\Delta h| \Delta h)=50$.}
\end{figure}

The accuracy of the heterodyned likelihood, even for LISA sources with signal-to-noise ratios in the thousands, such as the system shown in Figure~\ref{fig:likeLISA}, is impressive: absolute errors of order 0.02 and fractional errors of order $10^{-8}$.

\subsection{Noise Marginalization}

\begin{figure}[htp]
\includegraphics[width=0.48\textwidth]{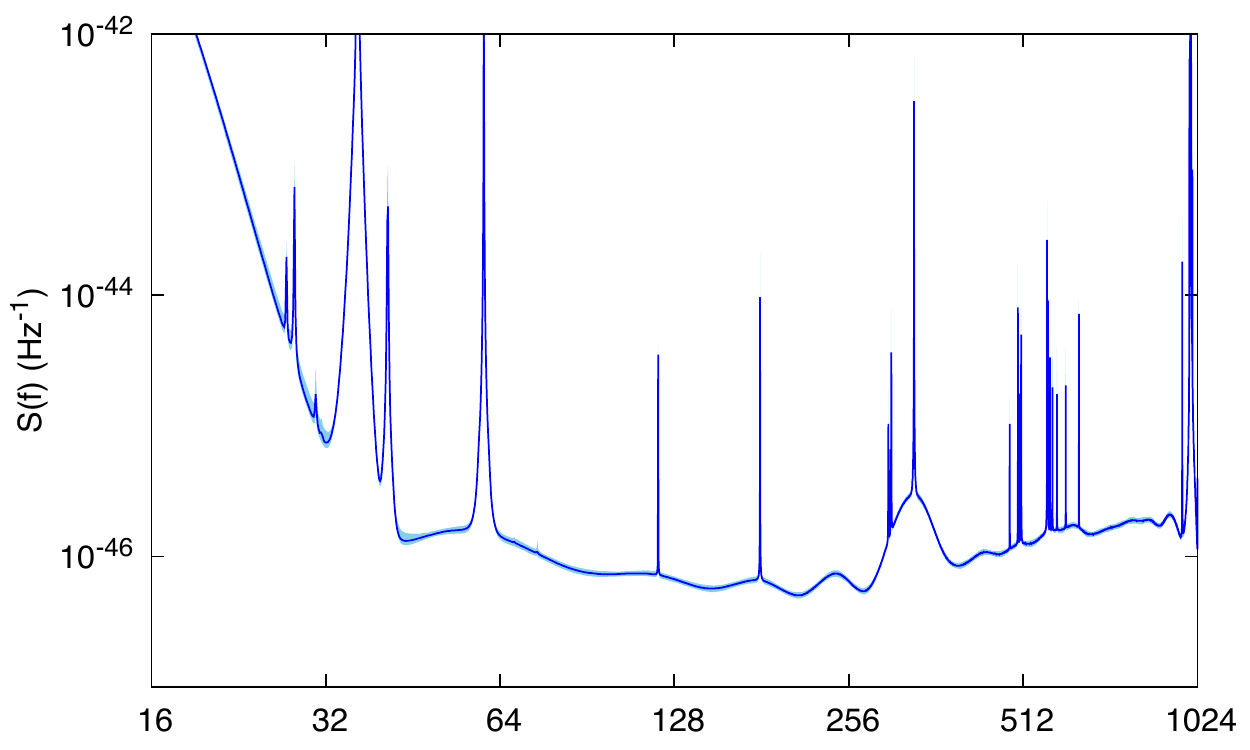} 
\caption{\label{fig:spec}  Power spectral density estimates for the LIGO Hanford detector using 4 seconds of data surrounding the binary merger GW151012. The light blue band indicates the 90\% credible region, while the solid dark blue lines indicates the median.}
\end{figure}

To produce parameter estimates that are robust against instrument calibration uncertainties and finite sample uncertainties in the on-source noise spectral estimates, it is desirable to marginalize over the calibration model and noise model. This poses a challenge for rapid parameter estimation techniques that rely on pre-computing quantities at full cadence. For the heterodyned likelihood, calibration uncertainties pose no problem as they introduce small changes in the amplitude and phase evolution that can be incorporated in the slow terms, and thus have minimal impact on the computational cost. In contrast, changes in the noise model are generally not smooth due to the presence of sharp spectral lines. Figure \ref{fig:spec} shows the median and 90\% credible band for the on-source power spectrum model in the LIGO Hanford detector using 8 seconds of data surrounding the black binary hole merger GW151012. Here the spectral model is a fixed dimension variant of the {\tt BayesLine} model~\cite{PhysRevD.91.084034} used by the {\tt QuickCBC}~\cite{Cornish:2021wxy} parameter estimation pipeline.

\begin{figure}[htp]
\includegraphics[width=0.48\textwidth]{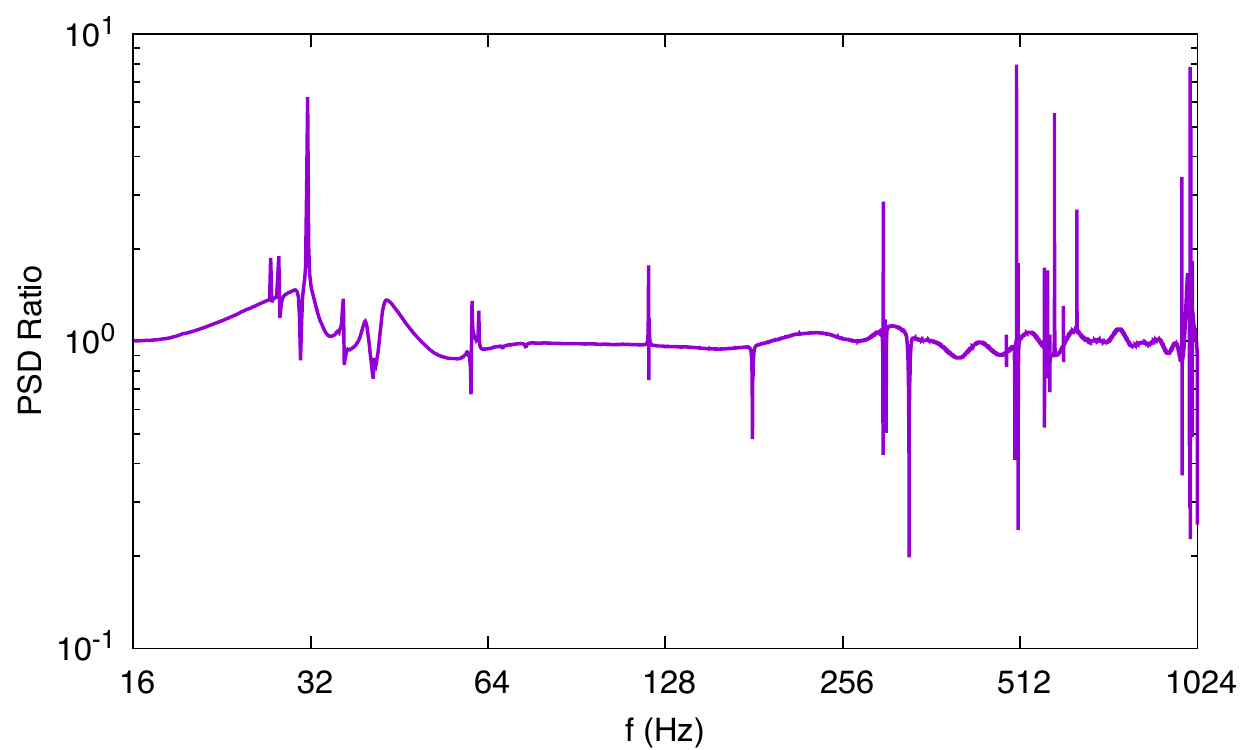} 
\caption{\label{fig:ratio}  The ratio of a fair draw from the power spectrum model and the reference power spectrum model for the LIGO Hanford detector using 8 seconds of data surrounding GW151012.}
\end{figure}

\begin{figure}[htp]
\includegraphics[width=0.48\textwidth]{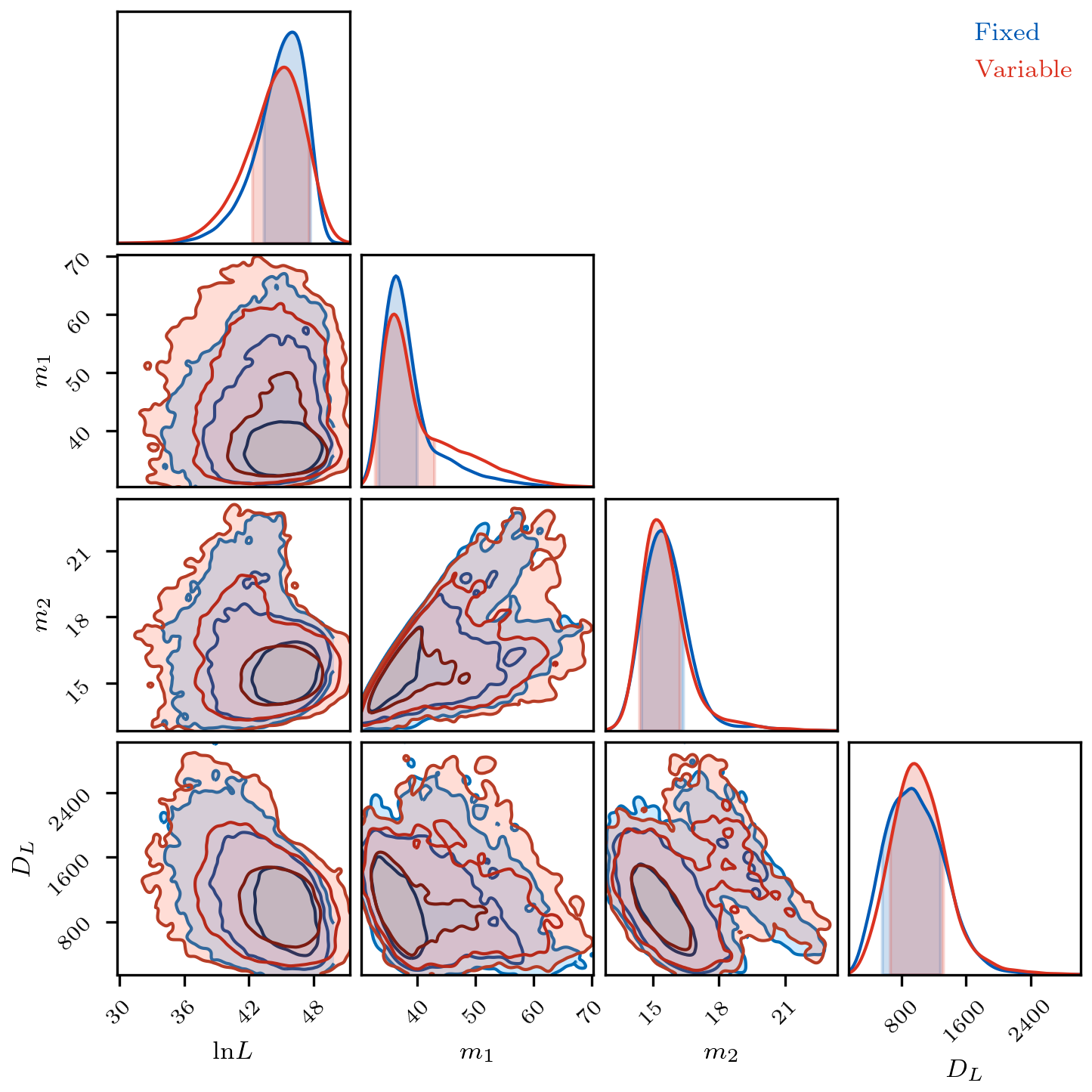} 
\caption{\label{fig:compare}  Posterior distributions for GW151012 with (Variable) and without (Fixed) marginalization over the power spectral density. Marginalizing over the PSD slightly inflates the spread in the posterior distributions.}
\end{figure}

The variations in the power spectrum look small when viewed on a logarithmic scale, and it is tempting to try and incorporate these variations by writing $\kappa(f) = d(f) \bar{h}^*(f)/\bar{S}(f)$ for the fast term and $\zeta(f) = \bar{S}(f) h(f)/ ( S(f) \bar{h}(f))$ for the slow term, where $\bar{S}(f)$ is some reference model for the PSD. Figure \ref{fig:compare} compares the ratio of a fair draw from the power spectrum to the median of the spectral model. The spectral lines lead to sharp features in the ratio that prevent it from being incorporated into the slow varying terms in the heterodyned likelihood. One way of handling the lines would be to excise the region around each line and calculate the likelihood directly in those regions and use the heterodyne for the remainder. A simpler approach is to use ``blocked Gibbs'' sampling, whereby the MCMC sampler alternates between updating the source parameters and the noise model parameters, with each noise model update followed by a re-computation of the Legendre polynomial expansion of the slow terms. This latter approach is relatively inexpensive since the reference waveform does not have to be recomputed, just the Legendre expansion. Typically, the cost of the re-computation is several times less than a standard likelihood evaluation. Moreover, if using multiple chains in a parallel tempered set-up, it is sufficient to limit the noise model updates to the cold chain, and to share the update with the hot chains. In this way, noise marginalization can be incorporated at little additional cost.

Figure \ref{fig:compare} compares the posterior distributions for the masses and distance of binary merger GW151012 with and without noise marginalization. Marginalizing over the noise model slightly inflates the widths of the posterior distributions, and slightly shifts the peaks. Similar small shifts were found for other systems in the LIGO-Virgo catalog~\cite{LIGOScientific:2018mvr,Abbott:2020niy}, suggesting that noise marginalization is probably less impactful than waveform uncertainties or calibration uncertainties (a similar conclusion was reached in Ref.~\cite{PhysRevD.102.023008}. 

The heterodyne procedure can also be applied to data with non-stationary noise that is locally stationary~\cite{dahlhaus1997,10.2307/41806549}.  Locally stationary data can be whitened and made stationary by transforming to the discrete wavelet domain, and dividing each wavelet pixel by the square root of the dynamic, or evolutionary pectrum, $S(f,t)$. This procedure de-correlates the noise in both time and frequency~\cite{Cornish:2020odn}. The rescaled white/stationary data can then be transformed to the frequency domain and used in the heterodyned likelihood. The steps are illustrated in equation (\ref{white}).
\begin{equation}\label{white}
d(t)  \mathrel{\mathop{\rightarrow}_{\mathrm{DFT}}} d(t,f) \mathrel{\mathop{\rightarrow}_{\mathrm{decorr}}}  \frac{d(t,f)}{\sqrt{S(f,t)} }  \mathrel{\mathop{\rightarrow}_{\mathrm{FT}}} \tilde{d}_w(f) \, .
\end{equation}
The waveform template $\tilde{h}(f)$ can be dynamically whitened using the time-frequency mapping 
\begin{equation}\label{tf}
t(f) = \frac{1}{2\pi} \frac{d \bar \Phi(f)}{df} \, ,
\end{equation}
and defining $S(f) = S(f,t(f))$. There will be a different time-frequency mapping for each harmonic. Using this procedure, non-stationary gravitational wave data can be analyzed just as quickly as stationary data.

\section{Summary}

The heterodyned likelihood~\cite{Cornish:2010kf} can be used to dramatically speed up gravitational wave parameter inference without sacrificing accuracy. The heterodyning procedure can be efficiently implemented using discrete Legendre polynomial expansions and a dynamic spacing of the frequency samples. The method can be applied to any waveform model and detector configuration, with the largest savings in computational cost occurring for low mass systems. The savings decrease as the number of harmonics in the waveform model increase, since each harmonic has to be be treated separately. The heterodyning approach can incorporate marginalization over calibration uncertainties and variations in the noise model.

\section*{Acknowledgments}
The author thanks Tyson Littenberg and Katerina Chatziioannou for discussions about noise marginalization, and Soichiro Morisaki for providing feedback on an earlier draft. This work was supported by NSF award PHY1912053 and NASA LISA foundation Science Grant 80NSSC19K0320, and used data obtained from the LISA Data Challenge (https://lisa-ldc.lal.in2p3.fr) and the Gravitational Wave Open Science Center (https://www.gw-openscience.org), a service of LIGO Laboratory, the LIGO Scientific Collaboration and the Virgo Collaboration. LIGO is funded by the U.S. National Science Foundation. Virgo is funded by the French Centre National de Recherche Scientifique (CNRS), the Italian Istituto Nazionale della Fisica Nucleare (INFN) and the Dutch Nikhef, with contributions by Polish and Hungarian institutes.

\bibliography{refs}

\end{document}